\documentclass[prb,aps,twocolumn,showpacs,superscriptaddress]{revtex4}
\usepackage{graphicx,amsmath,amsfonts}

\def\eq{&=&} 

\begin{document}
\title{Formation of defects in multirow Wigner crystals}

\author{Alexios D. Klironomos}

\affiliation{American Physical Society, 1 Research Road, Ridge, NY 11961-9000}

\author{Julia S. Meyer}

\affiliation{SPSMS, UMR-E CEA / UJF-Grenoble 1, INAC, Grenoble, F-38054, France}

\date{\today}


\begin{abstract}
 We study the structural properties of a quasi-one-dimensional classical Wigner
 crystal, confined in the transverse direction by a parabolic potential. With
 increasing density, the one-dimensional crystal first splits into a zigzag
 crystal before progressively more rows appear. While up to four rows the
 ground state possesses a regular structure, five-row crystals exhibit defects
 in a certain density regime. We identify two phases with different types of
 defects. Furthermore, using a simplified model, we show that beyond nine rows
 no stable regular structures exist.
\end{abstract}

\maketitle

\section{Introduction}

The electron crystal has created considerable interest since its
possible existence was first pointed out by Wigner~\cite{Wigner}.
The three-dimensional Wigner crystal and its two-dimensional
counterpart have been extensively studied, and there exist beautiful
experimental realizations of the latter using electrons trapped on
the surface of liquid
Helium~\cite{Helium1,Helium2,Helium1.5,q1dhel}. More recently,
Wigner crystallization in one dimension has received renewed
interest~\cite{Matveev,KMM,MML,HB,TK,Bockrath,Pepper_1,Pepper_2,Eggert,MAP,InSb,Akhanjee,Fiete,Shulenburger,Piacente2,Piacente,CH,q1dhel};
for recent reviews see Refs.~\onlinecite{review} and
\onlinecite{review_2}.

The realization of a one-dimensional system requires the dominance
of the confining potential over internal energies, in particular,
the inter-particle interactions. Upon increasing density (and, thus,
the interaction energy), or weakening the confining potential, the
crystal deviates from its strictly one-dimensional structure. It has
been shown that at a critical density, a transition to a zigzag crystal
takes place~\cite{zigzag1,zigzag2,Piacente,MML,review}. Though not
for electrons, this zigzag transition has indeed been observed using
$^{24}$Mg$^+$ ions in a quadrupole storage ring~\cite{ions}.

Here we investigate the structural properties of the classical
quasi-one-dimensional Wigner crystal beyond the zigzag regime. While
previous investigations~\cite{Piacente} concentrated on regular
structures, we are interested in the formation of defects. From
symmetry considerations the assumption of regular crystals is
plausible at low densities when the number of rows is small,
however, its validity is not at all obvious once the lateral extent
of the crystal increases at higher densities. In fact, one expects a
nonuniform charge density in the direction transverse to the wire
axis. In particular, considering the electrostatics problem of
charges confined by a parabolic potential, $V(y)\propto y^2$, the
density profile should obey $n(y) \propto \sqrt{w^2-y^2}$, where $w$
is the width of the system~\cite{Larkin-q1D,Chklovskii-q1D}.
Therefore, the assumption of perfect rows with equal linear
densities should eventually break down. The formation of defects is
of particular interest because they will have a direct impact on the
transport properties of the system: while regular rows are locked,
defects are expected to be mobile.

\section{Model}

We consider classical particles in two dimensions interacting via
long-range Coulomb interaction. The system is assumed to be infinite
in the $x$-direction and confined in the transverse $y$-direction by
a parabolic confining potential $V_{\rm conf}$. The energy of the
system then reads
\begin{eqnarray}
{\cal H}\eq{\cal H}_{\rm int}+{\cal H}_{\rm conf}\\
\eq\frac{e^2}{2\epsilon}\sum_{i\neq j}\frac1{|{\bf r}_i-{\bf r}_j|}
+\frac12m\Omega^2\sum_iy_i^2,
\end{eqnarray}
where $\epsilon$ is the dielectric constant of the material and
$\Omega$ is the frequency of harmonic oscillations in the confining
potential.

At low densities, the system is one-dimensional, and the particles
minimize their mutual Coulomb repulsion by occupying equidistant
positions along the wire, forming a structure with short-range
crystalline order--the so-called one-dimensional Wigner
crystal~\cite{Wigner}. Upon increasing the density, the
inter-electron distance diminishes, and the resulting stronger
electron repulsion eventually overcomes the confining potential
$V_{\rm conf}$, transforming the classical one-dimensional Wigner
crystal into a staggered (zigzag) chain. From the comparison of the
Coulomb interaction energy $V_{\rm int}(r)=e^{2}/\epsilon r$ with
the confining potential an important characteristic length scale
emerges. Indeed, the transition from the one-dimensional Wigner
crystal to the zigzag chain is expected to take
place when distances between electrons are of
the order of the scale $r_0$ such that $V_{\rm conf}(r_{0})= V_{\rm
int}(r_{0})$. Within our model, {\em i.e.}, for a parabolic
confining potential and Coulomb interactions, the characteristic
length scale $r_0$ is given as
\begin{equation}
r_{0}=\left(2e^2/\epsilon m\Omega^2\right)^{1/3}.
\end{equation}
It is convenient for the following discussion to measure lengths in
units of $r_0$. To that purpose we introduce a dimensionless density
\begin{equation}
\nu=n_er_{0},
\end{equation}
where $n_e=N/L$ is the linear density of the system. Rescaling
lengths, the energy can be written as
\begin{eqnarray}
{\cal H}\eq E_0\left[\frac12\sum_{i\neq j} \frac1{|\hat{\bf
r}_i-\hat{\bf r}_j|}+\sum_i\hat y_i^2\right],
\end{eqnarray}
where $E_0=\left(e^4m\Omega^2/2\epsilon^2\right)^{1/3}$.

\begin{figure}[ht]
\centering
\includegraphics[width=3.2in]{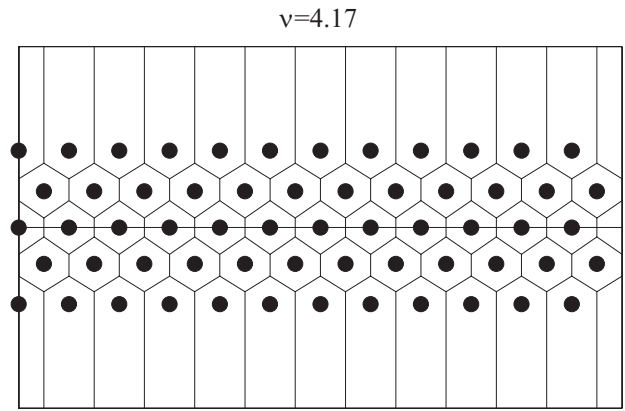}
\caption{\label{fig-reg_5}Regular structure with 5 rows at
$\nu=4.17$, shown with its Voronoi construction for illustration
purposes. This structure was obtained for 60 electrons in the unit
cell.}
\end{figure}

As a first step, we minimize the energy assuming regular rows,
aiming to find approximate values for the density range in which a
configuration with a given number of rows is stable. Assuming
staggering in the $x$-direction between neighboring rows and
inversion symmetry of the $y$-positions of the rows with respect to
the wire axis, the number of minimization parameters is $M/2$
($(M-1)/2$) for even (odd) number of rows $M$, and the minimization
is straightforward. Within these constraints, the minimization of
the energy with respect to the electron configuration
reveals~\cite{Piacente,zigzag1,KMM} that a one-dimensional crystal
is stable for densities $\nu<0.78$, whereas a zigzag chain forms at
intermediate densities $0.78<\nu<1.71$. More rows appear as the
density further increases. The number of rows as a function of $\nu$
is shown in Table \ref{table_1}.
\begin{table}[t]
\caption{\label{table_1}Number of rows in the crystal as a function
of the dimensionless density $\nu$, assuming regular structures.}
\begin{tabular}{|c|c|}
\hline
\# of rows ($M$)& density range\\
\hline
1 & $\nu<0.78$\\
2 & $0.78<\nu<1.71$\\
4 & $1.71<\nu<1.79$\\
3 & $1.79<\nu<2.72$\\
4 & $2.72<\nu<3.75$\\
5 & $3.75<\nu<4.84$\\
6 & $4.84<\nu<5.99$\\
\hline
\end{tabular}
\end{table}
One notices that, with the exception of the four-row
structure~\cite{Piacente} in the regime $1.71<\nu<1.79$, the linear
density per row $\nu_{\rm row}=\nu/M$ is of order $\lesssim 1$ in
all cases, \textit{i.e.}, another row is added to the crystal when
the distance between particles within a row is of the order of
$r_0$. A typical regular structure is shown in Fig. \ref{fig-reg_5}.

To investigate the importance of defects, the above conditions have
to be relaxed. In the following, we concentrate on the density
regime $1.79<\nu<5.99$, encompassing structures with 3 to 6 rows. In
Sec.~\ref{sec-method} the numerical method is introduced, and in
Sec.~\ref{sec-results} we present our results. In
Sec.~\ref{sec-simple} we introduce a simplified minimization
procedure that allows us to extend the calculation to a larger
number of rows, before concluding in Sec~\ref{sec-conclusion}.

\section{Numerical method}
\label{sec-method}

In order to find the ground state configuration of the system, the
energy of the electrons in the parabolic confining potential is
minimized with respect to the positions of the electrons for given
confinement strength and density. As the number of particles used in
the simulation is finite, commensurability effects are important. To
realize a regular $M$-row structure, the number of particles in the
simulation box has to be a multiple of $M$. Similarly, to realize a
defected structure, the defect density is determined by the number
of particles used in the simulation. To illustrate this, let us
consider a five-row structure. Regular structures are realized for
$N=5n$; for all other $N$, defects appear. As we expect the density
to be maximal at the center and decrease towards the edges, the
simplest symmetric defected structure possible is one where the
outer rows are missing one particle each compared to the inner rows,
\textit{i.e.}, structures of the form $[(n\!-\!1) n n n (n\!-\!1)]$.
Such structures are realized for $N=5n-2$. The defect density may be
defined as the number of missing particles in the outer rows divided
by the number of particles in the inner rows, $n_{\rm def}=(n_{\rm
inner}-n_{\rm outer})/n_{\rm inner}=1/n=5/(N+2)$. The minimum defect
density that can be realized is, therefore, determined by the
maximal number of particles that can be simulated. Thus, to find the
ground state of the system, we have to vary $N$ at fixed confinement
strength and density.

Conceptually, the proposed calculation is straightforward.  The
computational difficulty arises from the complexity of the
minimization problem. It is well known from the study of related
problems, e.g., the determination of the ground state of atomic
clusters or the optimal arrangement of charges in a two-dimensional
confined geometry~\cite{gen_alg_1,gen_alg_2,gen_alg_3}, that the
corresponding energy functional has a number of metastable states
that increases exponentially with the number of particles. In such a
case, classic minimization techniques are not the optimal choice.

Hybrid techniques employing genetic algorithms have been used in
many related problems~\cite{gen_alg_1,gen_alg_2,gen_alg_3} as a
general tool to explore the available phase space more thoroughly
and obtain better solutions with comparable computational cost to
conventional optimization techniques. One frequently finds that
counterintuitive disordered structures are favored.

For our case, a simulation box of finite length $L$ containing $N$
electrons is used. Periodic boundary conditions in the $x$-direction
are enforced to remove size effects. For the summation of the
interaction series, a quasi-one-dimensional restriction of the Ewald
method is employed, following a similar technique to that reported
in Ref.~\onlinecite{Ewald}. The appropriate methods of proven
stability for our quasi-one-dimensional geometry are of complexity
$O(N^2)$ and this fact, in conjunction with the significant number
of minimizations that need to be carried out (various system sizes
for given total linear density), implies the necessity of
substantial computational resources.

The total energy per particle of a particular configuration of $N$
electrons $\{r_{ij}\}$ can be written as
\begin{equation}
E\left[\{r_{ij}\}\right]=\frac{E_0}N
\left\{\frac{r_0}{L}\varepsilon\left[\{r_{ij}\}\right] +
\left(\frac{L}{r_0}\right)^{2}\sum_{ij}y_{ij}^2\right\},
\end{equation}
where $E_0$ is the previously defined energy scale and distances are
now measured in units of $L$. The complicated expression for the
(dimensionless) interaction energy $\varepsilon$ and the details of
its calculation are shown in appendix \ref{ewald_appendix}. For a
given number of electrons $N$ in a cell of length $L$, and a given
density $\nu$, one has to minimize $E\left[\{r_{ij}\}\right]$ with
respect to the electron configuration and thereby obtain the stable
structure with energy $E_{\rm GS}(\nu;N)$.

In a nutshell, the algorithm proceeds along the following steps: An
initial population of structures with random arrangements of
electrons within the cell is partially relaxed towards a (local)
minimum by a small number of iterations of a conventional
minimization algorithm. Every member of the original population is
then randomly split into two pieces, and the next generation is
created by merging the pieces in all possible combinations while
conserving the total number of particles. Subsequently, all newly
obtained structures are fully relaxed to a (perhaps only local)
minimum by a conventional minimization algorithm. A number of them
is then chosen as parent structures for the next generation, always
maintaining an appropriate diversity in the available
configurations, \textit{i.e.}, a wide enough distribution in
energies. The structure with the minimum energy is always retained
to serve as a parent. The entire cycle is repeated until acceptable
convergence is achieved. As expected, this hybrid approach is
superior to simple minimization: it rapidly and consistently
converges to complicated structures, avoiding being trapped in local
minima.

In the end, to find the ground state configuration of the system at
a given density $\nu$, the structure with the lowest energy, $E_{\rm
GS}=\min_N\{E_{\rm GS}(\nu;N)\}$, is chosen.

\section{Results}
\label{sec-results}

With the method described above, we are able to consider systems
comprised of up to $N\sim200$ electrons in the unit cell. We find
that the lowest energy structures for a given energy are either
regular structures, or structures where the linear density of the
outer-most rows, $\nu_{\rm outer}$, is lower than the linear density
of the inner rows, $\nu_{\rm inner}$~\footnote{Structures with defects
in inner rows do appear. We find, however, that they always have
higher energy.}. The finite number of particles in the unit cell
implies a lower limit to the defect density we can consider. Here we
define the defect density as $n_{\rm def}=1-\nu_{\rm outer}/\nu_{\rm
inner}$. For up to 6 rows, the number of particles per row exceeds
30. We are, therefore, able to identify defected structures
with linear defect densities $n_{\rm def}$ down to $\sim0.03$.

\begin{figure}[t]
\centering
\includegraphics[width=3.2in]{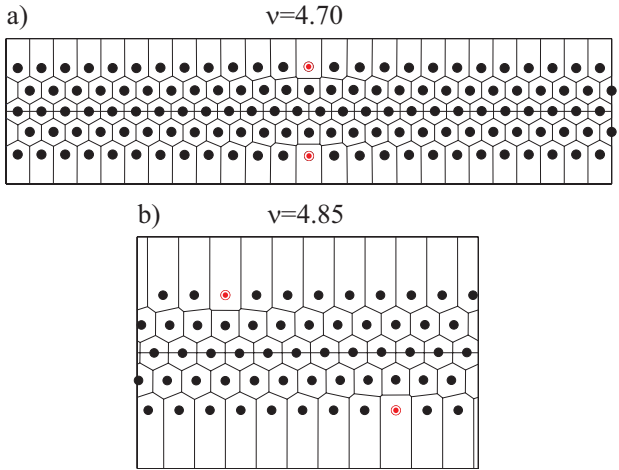}
\caption{\label{fig-def} Defected structures with 5 rows at a)
$\nu=4.70$ and b) $4.85$. The unit cell consists of 128 and 58
electrons, respectively, with the two outer rows missing an
electron. The corresponding defect densities are $n_{\rm
def}^{4.70}=1/26\approx0.038$ and $n_{\rm
def}^{4.85}=1/12\approx0.083$. Electrons in red half-filled disks
indicate the formal centers of the defects encountered.}
\end{figure}

Let us summarize our main findings before discussing them in more
detail: Up to 4 rows, the ground state of the system is free of
defects. In the five-row structure, defects appear as one approaches
the transition to 6 rows. Typical examples of such defected
structures are shown in Fig.~\ref{fig-def}. We find that the defect
density quickly increases with density and then levels off at values
of the order $n_{\rm def}\sim 0.08$. Note that different types of
defects appear: In the low density regime where the defect density
rapidly increases with density, the structure possesses inversion
symmetry with respect to $x$-axis, \textit{i.e.}, the centers of the
defects in the two outer rows are located at the same $x$-position
as shown in Fig.~\ref{fig-def}a. By contrast, the structures with
the maximal defect density $n_{\rm def}\sim 0.08$ display defects
that are maximally shifted with respect to each other along the
$x$-direction as shown in Fig.~\ref{fig-def}b. The transition to
six-row structures is shifted to a larger density as compared to the
value given in Table~\ref{table_1}. Above the transition, the ground
state is a regular six-row structure. Only upon further increasing
the density do defects appear again, before the transition to a
seven-row structure. Further analyzing the spatial structure of the
ground state configurations, we find that the presence of defects in
the outer rows also affects the particle positions in the inner
rows. While structures without defects consist of straight rows
without corrugation, structures with defects display corrugation,
\textit{i.e.}, distortions of the regular structure in both $x$- and
$y$-direction.

\subsection{Defects in five- and six-row crystals}

Using the full numerical minimization procedure, we find that the
five-row Wigner crystal is stable in the density range $3.75 < \nu <
4.86$. Defected structures replace the regular ground state at
$\nu_c^{(5)}=4.695$ and persist until the transition to 6 rows. Note
that, as the finite number of particles limits the defect densities
we can probe, the value $\nu_c^{(5)}$ represents an upper boundary
for the range of stability of the regular structure.

\begin{figure}[t]
\includegraphics[width=3.4in]{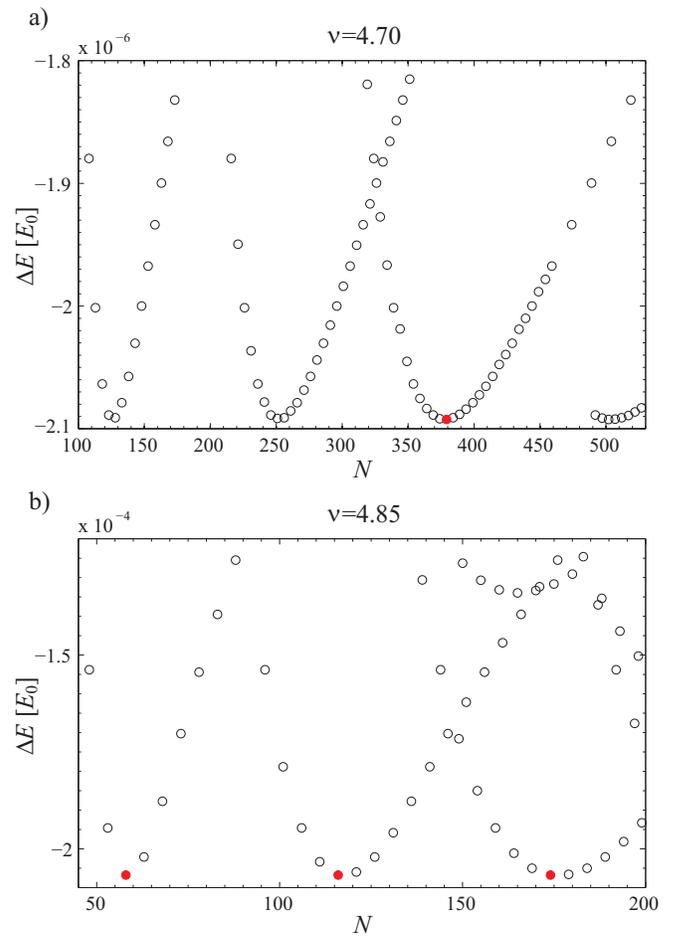}
\caption{\label{fig-E0} $E_{\rm GS}(N)$ at fixed density for a)
$\nu=4.70$ and b) $\nu=4.85$. The energy is measured in units of
$E_0=(e^4m\Omega^2/2\epsilon^2)^{1/3}$ from the lowest-energy
regular structure. a) For $\nu=4.70$, four minima can be clearly
seen. As the defect densities sampled differ slightly, they are not
exactly equal in energy. The global minimum is found at $N=379$,
corresponding to a defect density $n_{\rm def}^{4.70}=0.390$. b) For
$\nu=4.85$, three degenerate minima appear at $N=58, 116, 174$, all
corresponding to a defect density of $n_{\rm def}^{4.85}=0.083$.}
\end{figure}
\begin{figure}[t]
\includegraphics[width=3.4in]{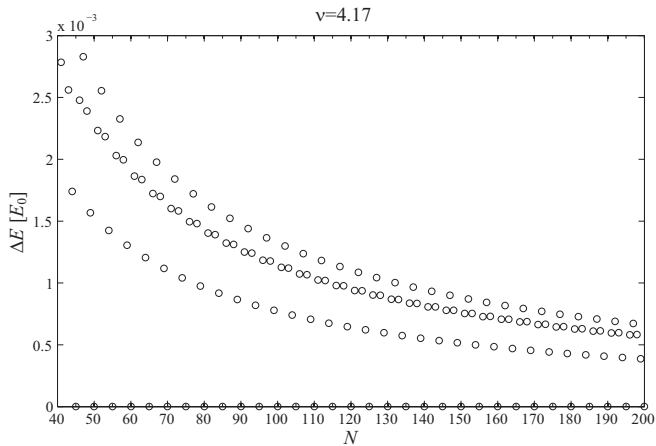}
\caption{\label{fig-E0_reg} Difference in energy at fixed density
$\nu=4.17$ as a function of the size of the unit cell.
$N=5n-1$ structures [$(n\!-\!1)nnnn$] form the lowest
excitation branch, followed by $N=5n+1$ [$n(n\!+\!1)nnn$],
$N=5n-2$ [$(n\!-\!1)nnn(n\!-\!1)$], and $N=5n-3$
[$(n\!-\!1)(n\!-\!1)nn(n\!-\!1)$]. Fitting the lowest excitation
branch to a general functional form $\alpha+\beta/N^\gamma$ we
obtain $\gamma\sim1$ and an energy gap in the thermodynamic limit
$\Delta E_{\infty}=2\times10^{-7}$.}
\end{figure}
\begin{figure}[t]
\includegraphics[width=3.4in]{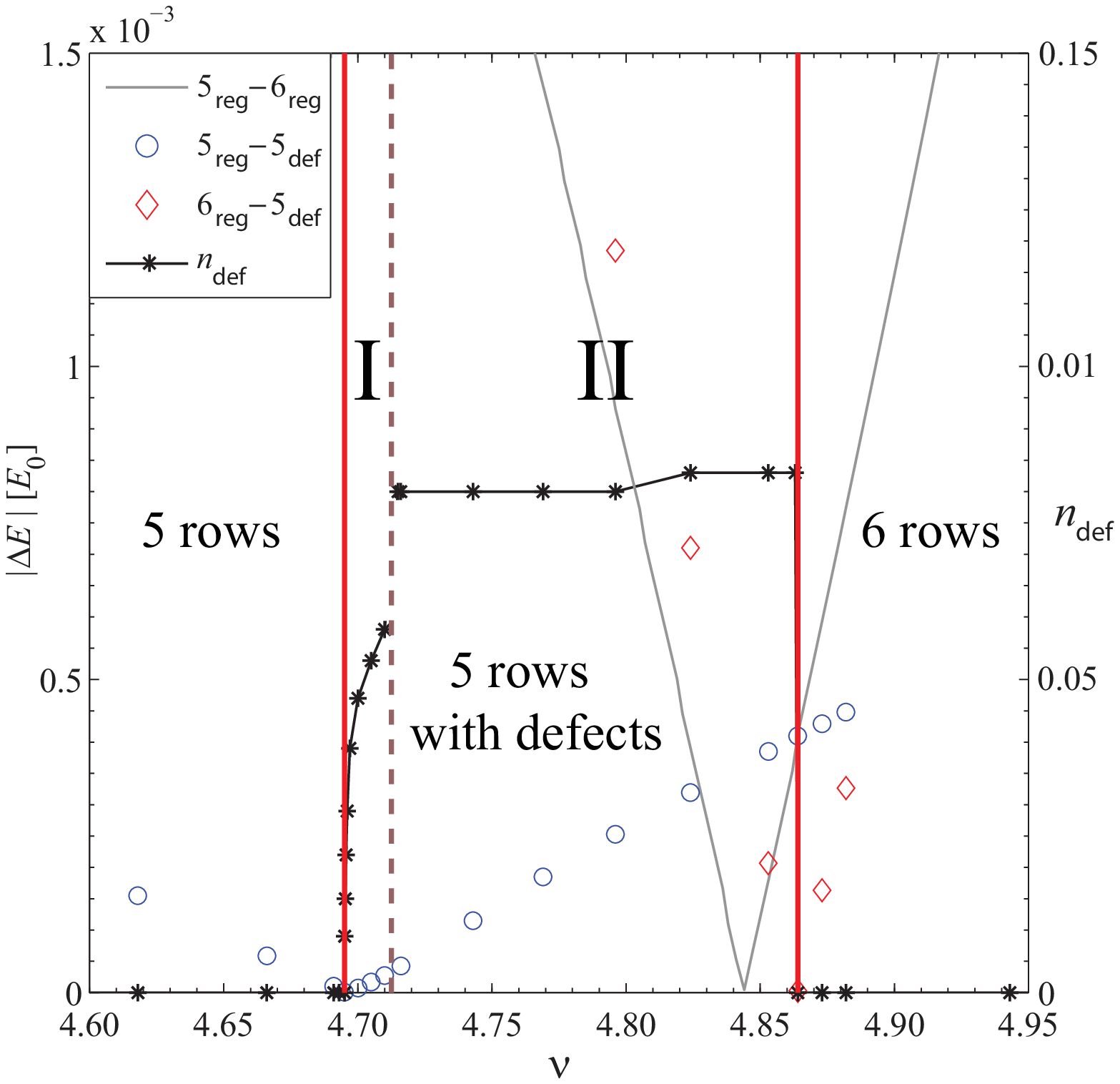}
\caption{\label{fig-nu_def} Energy gaps $|\Delta E|$ and boundaries
for the various phases encountered. The red vertical lines show the
phase boundaries as obtained by the location of the zero of the
corresponding energy difference between: five-row regular and
five-row defected structures (blue circles), six-row regular and
five-row defected structures (red rhombi). The dashed line indicates
the location of a first order transition between the two types of
defects encountered: in region I the centers of the defects coincide
whereas in region II the centers of the defects are maximally
separated within the unit cell. The grey line shows the energy
difference between five- and six-row regular structures. (Its zero
is the prediction for the phase boundary under the assumption of
regular rows.) Furthermore, the black stars show the defect density
$n_{\rm def}$ as a function of density $\nu$. Note the jump at the
boundary between regions I and II.}
\end{figure}

A regular five-row crystal is shown in Fig.~\ref{fig-reg_5} whereas
two defected five-row crystals are shown in Fig.~\ref{fig-def}. The
five-row crystals in Fig.~\ref{fig-def} correspond to $\nu=4.70$ and
$\nu=4.85$, and have a defect density $n_{\rm def}=0.038$ and
$n_{\rm def}=0.083$, respectively. Fig.~\ref{fig-E0} shows how the
defected structures were identified. In particular, the ground state
energy at fixed density $\nu=4.85$ as a function of the number of
particles in the unit cell is shown. Regular five-row structures are
realized for $N=5n$ with $n\in\mathbb{N}$. For all other $N$,
defected structures are obtained. Let us discuss the high density
structure $\nu=4.85$ displayed in Fig.~\ref{fig-E0}b first. Three
equivalent minima at $N=58$, $N=116$, and $N=174$ can be clearly
seen. These minima correspond to configurations with defects in the
outer-most rows, \textit{i.e.}, the outer rows have less particles
than the inner rows, namely $N_{\rm inner}=12 \; (24, 36)$, and
$N_{\rm outer}=11 \; (22, 33)$, respectively. The corresponding
defect density is $n_{\rm def}^{4.85}=1-11/12=0.083$. The low-energy
structure $\nu=4.70$ displayed in Fig.~\ref{fig-E0}a displays only
one minimum at $N=128$ within the regime that can be explored by the
full minimization procedure. This minimum corresponds to a defect
density $\tilde n_{\rm def}^{4.70}=0.0385$. To rule out finite size
effects, we extended our calculation to a larger number of particles
employing conventional minimization techniques, utilizing as
starting guesses the structures obtained form the full minimization
in the smaller unit cell. As Fig.~\ref{fig-E0}a shows, further
minima appear at $N=251$, $N=379$, and $N=502$ corresponding to
approximately the same defect density. In fact, the lowest energy
structure is obtained for $N=379$ where $n_{\rm
def}^{4.70}=0.0390$.

For comparison, Fig.~\ref{fig-E0_reg} shows the equivalent diagram
at a lower fixed density $\nu=4.17$ corresponding to the regular
ground state shown in Fig.~\ref{fig-reg_5}. The energy of defected
structures keeps decreasing with defect density until the lowest
defect density $n_{\rm def}=1-39/40=0.025$ reached given our
limitation on the number of particles. Note that the lowest
excitation branch shown in the picture corresponds to structures
missing one particle from only one of the outer rows. Fitting that
branch to a general functional form $\alpha+\beta/N^\gamma$ we
obtain $\gamma\sim1$ and an energy gap in the thermodynamic limit
$\Delta E_{\infty}=2\times10^{-7}$. Thus, we do not expect that the
regular structure becomes unstable at even lower defect densities.

Our findings in the vicinity of the transition from 5 to 6 rows are
summarized in Fig.~\ref{fig-nu_def}. The defect density as well as
the energy gaps to the lowest-lying regular or defected structure
are shown as a function of density. Note that due to the substantial
computational effort involved, the density interval is not uniformly
sampled. As mentioned earlier, the defect density quickly increases
in a narrow density interval and then levels off to an almost
constant value until the transition to six rows is reached. The
six-row crystal is  stable in the density range $4.86 < \nu < 6.04$.
It develops defects at around $\nu=5.75$, which also persist until
the transition to 7 rows.

To better understand the structures that appear we now turn to a
more detailed analysis of the spatial arrangement of particles in
the crystal.

\subsection{Analysis of row corrugation}

As can be seen from Fig.~\ref{fig-def}, two types of defected
structures appear. These two structures can be distinguished by
analyzing the distortion of the crystal. The distances between rows
vary as a function of density. While regular structures consist of
straight rows, structures with defects display corrugation. Let us
label the positions of particles as ${\bf r}_j^{(k)}$, where $j$
denotes the row and $k$ denotes the position along the row. In
regular structures, we find $y_j^{(k)} = y_j^{(k')}$ for all
$j,k,k'$, within the accuracy of the calculation. For the defected
structures, we define the average displacement of each row
\begin{figure}[t]
\includegraphics[width=3.4in]{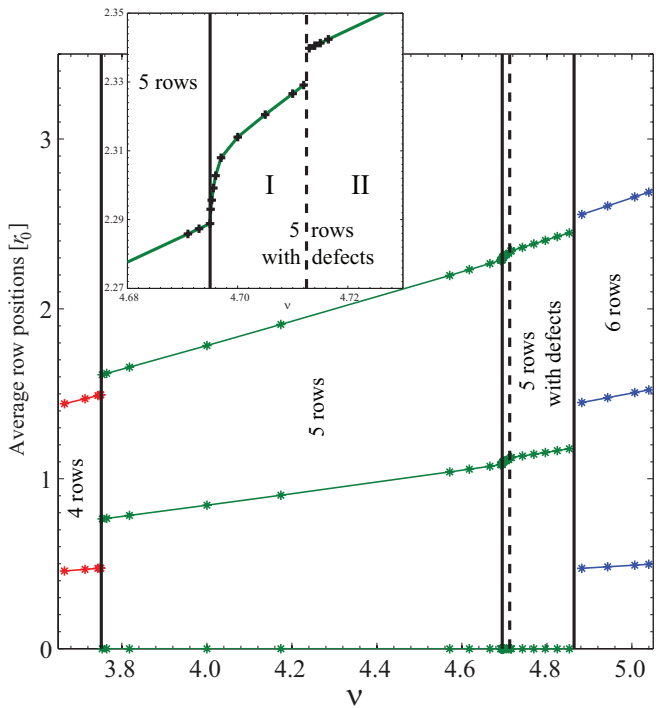}
\caption{\label{fig-y} Average positions of the crystal rows in
units of $r_0$ as a function of density. Note that due to symmetry,
only half of the rows are shown. The inset shows the detailed
behavior at the transition from the regular five-row to the defected
five-row structure. For regular structures, the distance between
rows increases linearly with $\nu$. In region I, the distance
between rows increases more rapidly. At the boundary between regions
I and II, there is a jump. In region II, the distance increases
again linearly with the same slope as for the regular structures.}
\end{figure}
\begin{eqnarray}
\overline{y}_j\eq\sum_{k=1}^{N_j}y_j^{(k)}/N_j,
\end{eqnarray}
where $N_j$ is the number of particles in row $j$. In
Fig.~\ref{fig-y}, the average positions of the rows are shown. Due
to the symmetry of the structure only half of them are displayed.

As expected, there are jumps at the transition to a structure with a
larger number of rows; in the intermediate region the distance grows
linearly with density. Interestingly, as shown in the inset, a
continuous transition to the structures with defects appears to take
place. The transition also marks the onset of corrugation in the
crystal structure. However, within the density regime of defected
structures, we find a discontinuity. In the region labeled I, the
distance between rows increases rapidly, a behavior that is well
fitted by a square root. At the boundary between regions I and II,
the row position displays a jump before it increases linearly again
in region II. This suggests two different defected phases that can
be characterized by their corrugation.

A close look at the defected structure reveals that the corrugation
exists in both directions, along and perpendicular to the wire axis.
We define the corrugation in the $y$-direction as the deviation from
the average row position in that direction,
\begin{figure}[t]
\includegraphics[width=3.4in]{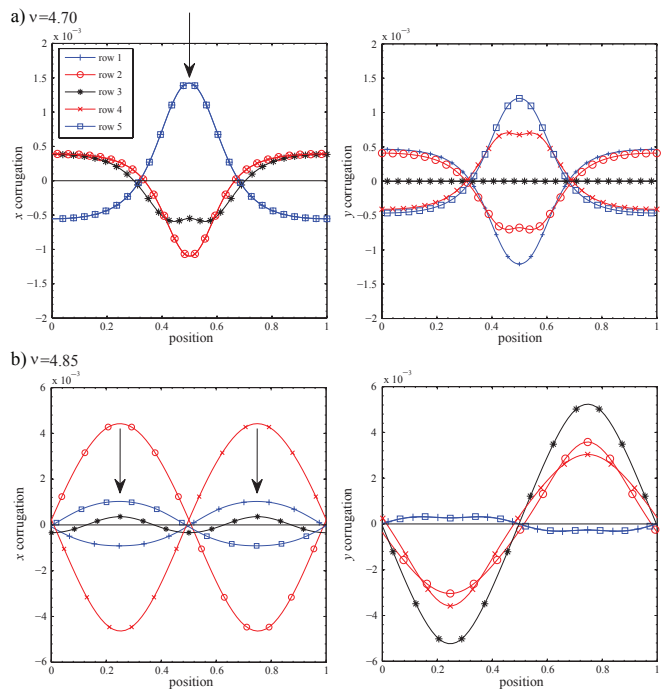}
\caption{\label{fig-delta_yx} Corrugation in the transverse (right
panels) and longitudinal (left panels) direction for the defected
structures that appear at a) $\nu=4.70$ (top panels) and b)
$\nu=4.85$ (bottom panels). For the definition of the corresponding
quantities please see the main text. The corrugation in the
longitudinal direction (left) is measured in units of the average
inter-particle distance within the row, while that in the
transverse direction (right) is measured in units of the average
distance between rows. The arrows indicate the particle located at
the center of the defect in each case. See also Fig.~\ref{fig-def}.}
\end{figure}
\begin{eqnarray}
\delta y_j^{(k)}\eq y_j^{(k)}-\overline{y}_j.
\end{eqnarray}
We can also define the average inter-particle distance for a given
row $j$ by
\begin{eqnarray}
\overline{\Delta x}_j\eq\sum_{k=1}^{N_j}(x_j^{(k+1)}-x_j^{(k)})
/N_j=\nu_j^{-1},
\end{eqnarray}
where $\nu_j$ is the dimensionless density in that row. Note that
$\nu=\sum_{n=1}^M\nu_j$. Subsequently, we define the corrugation in
the $x$-direction by
\begin{eqnarray}
\delta(\Delta x)_j^{(k+1\,k)}\eq
x_j^{(k+1)}-x_j^{(k)}-\overline{\Delta x}_j.
\end{eqnarray}

While the corrugation is less than one percent in both directions,
it turns out to be very important in determining the ground state of
the system. Figure~\ref{fig-delta_yx} shows examples of the two
dominant types of corrugation accompanying five-row defected
structures. Note that the arrows indicate the particle located at
the center of the defect in each case. As before, the chosen density
values are $\nu=4.70$ and $\nu=4.85$, close to the boundaries shown
in Figs.~\ref{fig-nu_def} and \ref{fig-y}. Fig.~\ref{fig-delta_yx}a
shows the corrugation for $\nu=4.70$, {\em i.e.}, a structure close
to the density where defects first appear. This kind of corrugation
is typical for the narrow density regime $4.695<\nu<4.712$, where
the defect density rapidly increases with $\nu$.
Fig.~\ref{fig-delta_yx}b shows the corrugation for $\nu=4.85$ with
defect density $n_{\rm def}=0.083$. This kind of corrugation is
characteristic of the structures exhibiting the maximum defect
density $n_{\rm def}\sim 0.08$.

Qualitatively, the two types of structures exhibit different
features. In the first defected structure that is encountered, see
Fig.~\ref{fig-delta_yx}a, the defects in the exterior rows are
rather localized, and they are located at the same position along
the crystal. The displacements are maximal for the outer rows and
decrease as one moves towards the interior of the crystal. In
particular, due to the symmetry of the defect, the innermost row
exhibits no corrugation in the $y$-direction at all.  At higher
density, both the $x$- and $y$-corrugations are approximately
sinusoidal. Furthermore, the defects on the two exterior rows are
maximally separated, \textit{i.e.}, they are shifted by half a
period, see Fig.~\ref{fig-delta_yx}b. For this kind of structures,
the center line possesses the maximum amplitude of $y$-oscillations.
A possible explanation is that, while the interaction energy (which
drives the corrugation) is only sensitive to the relative
corrugation, a deformation of the inner rows entails a smaller
change in confining potential energy.

We, thus, encounter two distinct phases with defects.
Fig.~\ref{fig-simple45}a shows the energy as a function of defect
density for different densities close to the boundary between the
two phases. Two minima corresponding to the different types of
defected structures can be clearly identified. The position of one
of the minima changes rapidly with density. This minimum corresponds
to the type of defect encountered in the low-density regime. The
position of the other minimum barely shifts with density. This
minimum corresponds to the sinusoidal defects encountered in the
high-density regime. At low density, it describes a metastable
state. However, its energy with respect to the other minimum
decreases with density until, at $\nu_{\rm I-II}^{(5)}=4.712$, it
eventually becomes the global minimum and, therefore, the ground
state. Both the defect density (Fig.~\ref{fig-nu_def}) and the
distance between rows (Fig.~\ref{fig-y}) display a discontinuity at
the transition. The transition between the two defected phases is,
thus, of first order.

The nature of the transition from regular to defected structures is
more difficult to identify as it requires going to very low defect
densities. The fact, that with decreasing density, the defect
density becomes lower and lower, until we reach the minimal value we
can simulate, suggests, however, that the transition might be second
order. In order to approach this transition, simplified models that
allow one to simulate a larger number of particles are required.
This models, then, may also be used in order to extend the
calculation to larger number of rows.

\section{Simplified minimization procedures}
\label{sec-simple}

The full minimization procedure is computationally intensive which
sets practical limits on the size of the unit cell one can simulate.
That in turn imposes constraints on the defect density. Therefore,
simplified models that allow us to simulate a larger number of
particles are worth investigating to gain a better understanding. We
start with the simplest model possible, compare with the
results of the full simulation described above, and then discuss
possible improvements.

Up to six rows, we find that defected structures have less particles
in the outer rows. In the previous section, we pointed out that
these defected structures display corrugation. As a first
approximation, one may neglect this corrugation and assume that all
rows are straight and regular, \textit{i.e.}, $ \delta
y_j^{(k)}=\delta(\Delta x)_j^{(k+1\;k)}=0$. Defects are incorporated
by allowing the linear densities in the inner and outer rows to
differ--in particular, the two outer rows have less particles,
$n_{\rm outer}<n_{\rm inner}$. The density of defects is controlled
by the parameter $\lambda=n_{\rm inner}/n_{\rm outer}$,
\textit{i.e.}, the density of defects is then given as $n_{\rm def}=
1-\lambda^{-1}$. In that case, for a fixed defect density, one has a
minimization involving $2M-1$ parameters, namely the $y$-position of
the rows and their relative shifts in the $x$-direction. Assuming
that defects are located in the outer rows, the calculation can be
further improved by \lq\lq unfreezing\rq\rq\ the $x$-positions of
particles in the outer rows. This is the method we will use in the
following. Given the much reduced parameter space, a conventional
minimization procedure is sufficient here.
\begin{figure}[t!]
\includegraphics[width=3.2in]{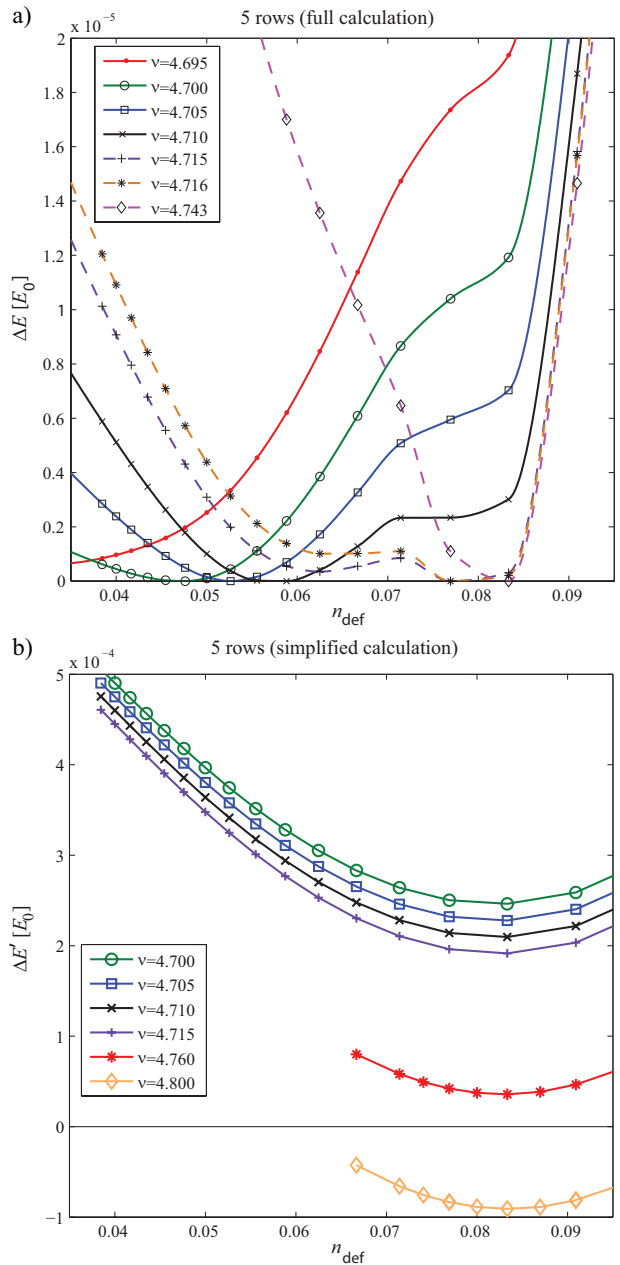}
\caption{\label{fig-simple45} a) Difference $\Delta E=E_{\rm
defect}-E_{\rm regular}$ for five-row structures as a function of
defect density $n_{\rm def}$ within a simplified calculation scheme
that incorporates defects by allowing the linear densities in the
inner and outer rows to be different. Note that there is a region
\textit{before} the critical $\nu=4.86$ of the $5\rightarrow 6$
transition where the defected structures become stable. b) Energy of
five-row defected structures at various densities as a function of
defect density $n_{\rm def}$ using the full minimization procedure.
The minima are shifted to zero for clarity of presentation. Note
that the minimum present at small defect densities is not captured
by the simplified method.}
\end{figure}

The findings for five-row structures are shown in
Fig.~\ref{fig-simple45}a. Note that the model captures correctly the
appearance of defects in the five-row structure, and it also
predicts a regular ground state for the four-row crystal.
Furthermore, the maximal defect density is reproduced: For the
five-row structure, we can see that the defect density $n_{\rm
def}\sim 0.08$ leading to the minimal energy is in agreement with
the full minimization.

Analyzing the results in more detail, (expected) discrepancies are
found. Due to the constraints imposed, the energies of defected
structures are too high. For the five-row structures for example,
the simplified model finds that defects appear around $\nu=4.78$
whereas the full minimization reveals that defected structures
become the ground state configuration already at a lower density
$\nu=4.695$. Furthermore, the simplified model does not capture the
rise in defect density up to the maximal value. In particular, the
simplified model completely misses the low-density regime with
symmetric defects. Comparing Figs.~\ref{fig-simple45}a and b, the
additional minimum at low-defect densities  present in
Fig.~\ref{fig-simple45}a is clearly absent in
Fig.~\ref{fig-simple45}b. Thus, the simplified method only captures
the high-density phase with sinusoidal defects. The reason is most
likely that it corresponds to a fairly smooth corrugation and,
therefore, is still present when one imposes straight rows. By
contrast, the minimum at low defect densities is associated with
fairly sharp features in the corrugation profile and may, therefore,
be suppressed by imposing straight regular rows. In particular, it
is straightforward to verify that, for constant linear density in
the inner rows, the defects on the outer rows  will be maximally
separated. In order to obtain a defected structure where the centers
of the defects coincide, a longitudinal distortion of the inner rows
is indispensable.

To summarize, the simplified model correctly reproduces the typical
defect density--though it overestimates the energy of the defected
structures which therefore are stable only in a reduced density
interval. However, the model does not reproduce the low-density
defected phase and, therefore, can not be used to explore the nature
of the transition from regular to defected structures. The method
may be used to study the stability of regular structures for
structures with more rows. We find that upon further increasing the
number of rows, the regime of densities where the ground state
contains defects widens. Under the assumption that only the outer
rows contain defects, regular structures disappear completely once
the number of rows exceeds nine, as is evident from table
\ref{table_2}. As the simplified method overestimates the energy of
defected structures and misses the phase with symmetric defects, it
is likely that regular crystals cease to be the ground state already
for a smaller number of rows.

As is also shown in Table \ref{table_2}, the typical defect density
increases with the number of rows and also slightly varies with
density for a given number of rows. Note that considering structures
of the type $[(n\!-1\!) n \dots n (n\!-1\!)]$, the defect densities
obtained can only take the discrete values $n_{\rm def}=1/n$.
\begin{table}[t]
\caption{\label{table_2}Number of rows and defect density in the
crystal as a function of the dimensionless density $\nu$. All
numbers shown were obtained using the simplified minimization
procedure described in Sec.~\ref{sec-simple}.}
\begin{tabular}{|c|c|c|c|}
\hline
\# of &total&density range&$n_{\rm def}$\\
rows&density range&with defects&\\
\hline
4 & $2.72<\nu<3.75$  &       N/A       & N/A\\
5 & $3.75<\nu<4.86$  &$4.77<\nu<4.86$  & 0.083\\
6 & $4.86<\nu<6.04$  &$5.75<\nu<6.04$  & 0.091--0.100\\
7 & $6.04<\nu<7.31$  &$6.76<\nu<7.31$  & 0.100--0.111\\
8 & $7.31<\nu<8.64$  &$7.80<\nu<8.64$  & 0.100--0.125\\
9 & $8.64<\nu<10.01$ &$8.87<\nu<10.01$ & 0.111--0.125\\
10& $10.01<\nu<11.37$&$10.01<\nu<11.37$& 0.125--0.143\\
11& $11.37<\nu<12.77$&$11.37<\nu<12.77$& 0.125--0.143\\
12& $12.77<\nu<14.19$&$12.77<\nu<14.19$& 0.143--0.167\\
13& $14.19<\nu<15.67$&$14.19<\nu<15.67$& 0.143--0.167\\
\hline
\end{tabular}
\end{table}

Eventually, one expects that more complicated structures will
appear. The simple configuration we studied is in competition with
structures where defects appear away from the edges, such as
structures of the type $[(n\!-\!1)(n\!-\!1) n \dots n
(n\!-\!1)(n\!-\!1)]$, for example. Detailed calculations within
these simplified models reveal that such structures are indeed
competitors for the ground state, but up to 13 rows such a minimum
is not realized.

To approach the transition between regular and defected rows, we use
a different trick. An unbiased search for the global minimum is
numerically costly because a simple minimization may get stuck in a
metastable minimum. However, if the initial guess of the electron
configuration is sufficiently close to the global minimum, a simple
minimization will converge. Having identified the structure of
defects in region I, one may feed such structures into a simple
minimization at lower densities. The results of such a procedure
have been included in Figs.~\ref{fig-nu_def} and \ref{fig-y}. There,
structures with defect densities down to $n_{\rm def} \sim 0.03$
were obtained with the full minimization, whereas structures close
to the phase boundary with lower defect structures were obtained
with the method described here. The results suggest that the defect
density indeed vanishes at the transition which points to a second
order phase transition.

\section{Conclusion}
\label{sec-conclusion}

We study quasi-one-dimensional systems of classical particles
interacting via long-range Coulomb interactions and confined by a
parabolic potential in the transverse direction. The ground state
configurations are multi-row Wigner crystals where the number of
rows is controlled by the density (or the strength of the confining
potential). We find that defects that accommodate the density
variation in the transverse direction appear once the number of rows
exceeds four.

Defected structures have less particles in the outer than in the
inner rows. The full numerical minimization for five rows reveals
that two distinct types of defected phases exist. Upon increasing
density, the regular structure at low-densities is replaced by a
structure with symmetric defects, \textit{i.e.}, where the center of
the defect on the two outer rows is located at the same
$x$-position. As the number of particles that can be simulated sets
a lower limit on the defect density that can obtained, the full
minimization allows one only to provide an upper limit for the
density at the transition from regular to defected structures. We
extend our calculations to lower densities by using structures with
the type of defect described above as the input for a simple
minimization. The results indicate that the defect density vanishes
at $\nu_c^{(5)}=4.695$ and that the transition is of second order.
To obtain symmetric defects, the longitudinal distortion of the
inner rows, namely an increased density at the center of the defect,
is crucial. Any analytical description of the transition would have
to take into account this distortion.

Upon further increasing density, the defect density rapidly
increases. At a critical density, $\nu_{\rm I-II}^{(5)}=4.712$,
structures with a different type of defect corresponding to a
sinusoidal distortion of the rows with a phase shift of half a
period between the two outer rows become become the ground state.
This second regime is characterized by a defect density that barely
varies with density and extends up to the transition to six rows.
The transition between the two defected phases with different
symmetries is first order.

Simplified models neglecting the corrugation of the rows only
capture this second defected phase. Thus, these models do not allow
one to further investigate the nature of the phase transition from
regular to defected structures. However, as this second phase
occupies most of the density interval, they may be used to study the
stability of regular structures upon increasing the number of rows.
We find that beyond nine rows, no stable regular structures exist.
Taking into account that the simplified model overestimates the
energy of defected structures, we expect that stable regular
structures may disappear even earlier.

\begin{acknowledgments}
We would like to acknowledge stimulating discussions with K.~A.
Matveev, Yu.~V.~Nazarov, and A.~Melikyan. Part of the calculations
were performed at the Ohio Supercomputer Center thanks to a grant of
computing time. This work was supported by the U.S. Department of
Energy, Office of Science, under Contract No.~DE-FG02-07ER46424.
\end{acknowledgments}

\appendix
\section{Ewald summation method for a quasi-one-dimensional geometry}
\label{ewald_appendix}

The method we use essentially follows the steps outlined in
Ref.~\onlinecite{Ewald}. It is based on the Poisson summation
formula relating summations over direct and reciprocal space,
\begin{equation}
\sum_{n=-\infty}^{+\infty}f(nL)=\frac{1}{L}\sum_{m=-\infty}^{+\infty}
F\left(\frac{2\pi m}{L}\right),
\end{equation}
where the Fourier transform of $f(x)$ is defined as
\begin{equation}
F(k)=\int_{-\infty}^{+\infty}\!\!\!\mathrm{d}x\,e^{ikx}f(x).
\end{equation}
Let us consider the function $f(x)=e^{-(\rho+x)^2t}$. By completing
the square and carrying out the Fourier transform integration, we
obtain the fundamental equation
\begin{equation}
\sum_{n=-\infty}^{+\infty}e^{-(\rho+nL)^2t}=
\frac{\sqrt{\pi}}{L}t^{-1/2}\sum_Ge^{iG\rho}e^{-\frac{G^2}{4t}},
\label{eq-Poisson}
\end{equation}
where the reciprocal lattice vectors are given by $G=2\pi m/L$ with
$m=0,\pm 1,\dots$. The following definition of the incomplete
$\Gamma$ function is extensively used and therefore given here for
reference:
\begin{equation}
\label{gam}
\frac{\Gamma({\mu,ux^2})}{x^{2\mu}}=
\int_{u}^{+\infty}\!\!\!\mathrm{d}t\,t^{\mu-1}e^{-x^2t}.
\end{equation}
The system we are considering contains a basic cell of length $L$
with $N$ electrons. The spatial extent in the $y$-direction is
limited by the confining potential. In the $x$-direction, we impose
periodic replications of the basic cell to avoid edge effects. The
interaction energy per cell of the system can be written
\begin{equation}
\tilde\epsilon[\{{\bf r}_{ij}\}]=\frac{1}{2}\sum_{i\ne j}
\sum_{n}\frac{q_iq_j}{|{{\bf r}}_{ij}+nL\hat{{\bf x}}|}
+\frac{1}{2}\sum_{j}q_j^2\sum_{n\ne 0}\frac{1}{|nL\hat{\bf x}|},
\end{equation}
where $q_i$ is the charge of particle $i$, ${\bf r}_{ij}={\bf
r}_i-{\bf r}_j$, and the index $n$ runs over replicas of the unit
cell. The artificial separation of the terms is for our convenience.
We then introduce the notation
\begin{align}
\label{phi}
\Phi({\bf r})&=\sum_n\frac{1}{|{\bf r}+nL\hat{\bf x}|}
\end{align}
for $|{\bf r}|\neq0$ and $\Phi_0=\sum_{n\ne
0}\displaystyle{1/|nL\hat{\bf x}|}$.

In what follows we will split the summations in direct and
reciprocal space. To cancel the divergencies appearing in the above
sums, we will assume a uniform neutralizing background charge.

Using Eq.~(\ref{gam}), we obtain the following representation,
\begin{align}
\Phi({\bf r})&=\frac{1}{\sqrt{\pi}}\sum_n
\int_0^{+\infty}\!\!\!\mathrm{d}t\,t^{-1/2}
e^{-|{\bf r}+nL\hat{\bf x}|^2t}.
\end{align}
Here we will introduce an artificial separation constant $\alpha$
which will control the splitting of the summation between direct and
reciprocal space. We then have $\Phi({\bf r})=\Phi^>({\bf
r})+\Phi^<({\bf r})$, where
\begin{align}
\Phi^>({\bf r})&=\frac{1}{\sqrt{\pi}}\sum_n
\int_{\alpha^2}^{+\infty}\!\!\!\mathrm{d}t\,t^{-1/2}
e^{-|{\bf r}+nL\hat{\bf x}|^2t}\notag\\
&=\sum_n\frac{\mathrm{erfc}(\alpha|{\bf r}+nL\hat{\bf x}|)}{|{\bf r}+nL\hat{\bf x}|}
\end{align}
and
\begin{align}
\Phi^<({\bf r})&=\frac{1}{\sqrt{\pi}}\sum_n
\int_0^{\alpha^2}\!\!\!\mathrm{d}t\,t^{-1/2}e^{-(x+nL)^2t}e^{-y^2t}.
\end{align}
To evaluate $\Phi^<({\bf r})$, we use Eq.~(\ref{eq-Poisson})
yielding
\begin{align}
\Phi^<({\bf r})&=\frac{1}{L}\sum_Ge^{iGx}
\int_0^{\alpha^2}\!\!\!\mathrm{d}t\,t^{-1}e^{-\frac{G^2}{4t}}e^{-y^2t}.
\end{align}
While for $G\neq0$ the integration yields incomplete Bessel
functions~\footnote{For an efficient method for the evaluation of
the incomplete Bessel function, we refer the reader to
Ref.~\onlinecite{harris}.},
\begin{eqnarray}
\int_0^{\alpha^2}\!\!\!\mathrm{d}t\,t^{-1}e^{-\frac{G^2}{4t}}e^{-y^2t}
=K_0\left(\frac{G^2}{4\alpha^2},\alpha^2y^2\right),
\end{eqnarray}
the $G=0$ term (denoted $I_0$ in the following) is divergent and has
to be treated separately. Using the substitution $z=\alpha^2/t$ and
expanding the second exponential, one finds
\begin{align}
I_0
&=\frac{1}{L}\lim_{G\rightarrow0}\int_{1}^{+\infty}\!\!\!\mathrm{d}z\,z^{-1}
e^{-\frac{G^2}{4\alpha^2}z}
\sum_{m=0}^{+\infty}\frac{(-1)^m}{m!}(\alpha y)^{2m}z^{-m}.\notag
\end{align}
The divergent contribution comes from $m=0$, namely
\begin{eqnarray}
\lim_{G\rightarrow0}\int_{1}^{+\infty}\!\!\!\mathrm{d}z\,z^{-1}
e^{-\frac{G^2}{4\alpha^2}z}\eq-\gamma+\ln4\alpha^2
-\lim_{G\rightarrow0}\ln G^2.\notag
\end{eqnarray}
The rest of the sum can be evaluated to
\begin{eqnarray}
&&\sum_{m=1}^{+\infty}
\int_{1}^{+\infty}\!\!\!\mathrm{d}z\,z^{-1}\frac{(-1)^m}{m!}(\alpha y)^{2m}z^{-m}\\
&=&\sum_{m=1}^{+\infty}
\frac{(-1)^m}{mm!}(\alpha y)^{2m}=-\gamma-\ln(\alpha^2 y^2)-\Gamma(0,\alpha^2 y^2).\notag
\end{eqnarray}
Thus,
\begin{align}
I_0&=-\frac{1}{L}
\left\{2\gamma+\lim_{G\rightarrow0}\ln G^2+\ln(y^2/4)
+\Gamma\left(0,\alpha^2y^2\right)\right\},\notag
\end{align}
and
\begin{eqnarray}
\Phi^<({\bf
r})&=&\frac1L\sum_{G\neq0}e^{iGx}
K_0\left(\frac{G^2}{4\alpha^2},\alpha^2y^2\right)+I_0.
\end{eqnarray}
Splitting up $\Phi_0$ in the same way, we find
\begin{align}
\Phi_0^>({\bf r})&=\sum_{n\neq0}
\frac{\mathrm{erfc}(\alpha|n|L)}{|n|L}
\end{align}
and
\begin{eqnarray}
\Phi_0^<({\bf r})&=&\frac{1}{L}\sum_G
\int_0^{\alpha^2}\!\!\!\mathrm{d}t\,t^{-1}e^{-\frac{G^2}{4t}}
-\frac1{\sqrt\pi}\int_0^{\alpha^2}\!\!\!\mathrm{d}t\,t^{-1/2}\notag\\
&=&\frac{1}{L}\sum_{G\neq0}\Gamma\left(0,\frac{G^2}{4\alpha^2}\right)\\
&&-\frac1L\left\{\gamma-\ln4\alpha^2
+\lim_{G\rightarrow0}\ln G^2\right\}-\frac{2\alpha}{\sqrt\pi}.\notag
\end{eqnarray}

At this stage we put everything together, $\tilde\epsilon[\{{\bf
r}_{ij}\}]=\frac12\sum_{i\neq j}q_iq_j\Phi({\bf
r}_{ij})+\frac12q_j^2\Phi_0$, and combining various terms we obtain
the result for the interaction energy per cell of the system,
\begin{widetext}
\begin{eqnarray}
\tilde\epsilon[\{{\bf r}_{ij}\}]&=&
\frac{1}{2}\sum_{i,j}q_iq_j
\left\{{\sum_n}'\frac{\mathrm{erfc}(\alpha|{\bf r}_{ij}+nL\hat{\bf x}|)}{|{\bf r}_{ij}+nL\hat{\bf x}|}
+\frac{1}{L}\sum_{G\ne 0}e^{iGx_{ij}}K_0\left(\frac{G^2}{4\alpha^2},\alpha^2y_{ij}^2\right)\right\}\\
&&-\frac{1}{2L}\sum_{i\ne j}q_iq_j\left[\gamma+\ln{\alpha^2y^2_{ij}}+\Gamma(0,\alpha^2y^2_{ij})\right]
-\frac{\alpha}{\sqrt{\pi}}\sum_jq_j^2
-\frac{1}{2L}\left(\sum_jq_j\right)^2\left[\gamma-\ln 4\alpha^2+\lim_{G\rightarrow0}\ln G^2\right],\notag
\end{eqnarray}
\end{widetext}
where the notation ${\sum_n}'$ implies that for $n=0$ there is no
self-interaction term in the summation. For a charge neutral system,
the last term vanishes. For a system of electrons, as the one under
consideration, a uniform positive neutralizing background will
exactly cancel the divergent term $\lim_{G\rightarrow0}\ln G^2$.

We define a dimensionless separation constant through
$\alpha=\widetilde\alpha/L$ and introduce dimensionless coordinates.
Subsequently, the dimensionless interaction energy per electron in
the simulation box can be cast as follows
\begin{eqnarray}
\varepsilon[\{{\bf r}_{ij}\}]=\frac{1}{2}\sum_{i,j}f[\{{\bf
r}_{ij}\}] -\frac{N\widetilde\alpha}{\sqrt{\pi}}
-N^2\left[\gamma-\ln{\left(\frac{2\widetilde\alpha}{L}\right)}\right],\notag\\
\end{eqnarray}
with
\begin{widetext}
\begin{eqnarray}
f[\{{\bf r}_{ij}\}]&=&{\sum_n}'
\frac{\mathrm{erfc}\left(\widetilde\alpha\sqrt{(x_{ij}+n)^2+y_{ij}^2}\right)}{\sqrt{(x_{ij}+n)^2+y_{ij}^2}}
+2\sum_{q=1}^{+\infty}\cos\left(2\pi qx_{ij}\right)
K_0\left(\frac{\pi^2q^2}{\widetilde\alpha^2},\widetilde\alpha^2y_{ij}^2\right)
-\ln\widetilde\alpha^2y_{ij}^2-\Gamma(0,\widetilde\alpha^2y_{ij}^2).
\end{eqnarray}
\end{widetext}

\end{document}